\documentclass[usenatbib,usegraphicx]{mn2e}
\usepackage{subfigure}
\usepackage{longtable}
\usepackage{graphicx}
\usepackage{graphics}
\usepackage{keyval}
\usepackage{trig}
\usepackage{graphicx}
\usepackage{dcolumn}
\usepackage{bm}
\def\beq{\begin{equation}}
\def\eeq{\end{equation}}
\def\msun{M_\odot}
\def\kms{\, {\rm km \, s}^{-1} }

\def\prd{Phys. Rev. D}
\def\mnras{MNRAS}
\def\apj{ApJ}

\def\apjl{ApJ}
\def\araa{ARAA}
	
\def\aap{A \& A}

\def\lcdm{{\Lambda}CDM}
\def\half{{1 \over 2}}
\def\aap{Astron. Astrophys.}
\begin{document}

\title{Are sterile neutrinos consistent with clusters, the CMB and MOND?}

\author{Garry W. Angus}
\author[G. W. Angus]{G. W. Angus$^{1,2}$\thanks{email:
gwa2@st-andrews.ac.uk}   \\
$^{1}$SUPA, School of Physics and Astronomy, University of St. Andrews, KY16 9SS Scotland\\
$^{2}$Dipartimento di Fisica Generale ''Amedeo Avogadro", Universit\'a degli studi di Torino, Via P. Giuria 1, I-10125, Torino, Italy }

\date{\today}
\maketitle
\begin{abstract}
If a single sterile neutrino exists such that $m_{\nu_s}\sim11eV$, it can serendipitously solve all outstanding issues of the Modified Newtonian Dynamics. With it one can explain the dark matter of galaxy clusters without influencing individual galaxies, match the angular power spectrum of the cosmic microwave background and potentially fit the matter power spectrum. This model is flat with $\Omega_{\nu_s}\sim0.23$ and the usual baryonic and dark energy components, thus the Universe has the same expansion history as the $\lcdm$ model and only differs at the galactic scale where the Modified Dynamics outperforms $\lcdm$ significantly.
\end{abstract}

\section{Introduction}
\protect\label{sec:intr}
Milgrom's Modified Newtonian Dynamics (MOND, see \citealt{milgrom83a,sanders02,bekenstein06,milgrom08}) as the weak acceleration limit of Einstein's general relativity is arguably consistent with a wide range of evidence from astronomical systems from the orbits of the planets in the solar system (where Newton's gravity suffices without dark matter) to dwarf galaxies (\citealt{milgrom95,angus08}) and globular clusters (Angus \& McGaugh 2008, in prep) of the Milky Way, tidal dwarf galaxies (\citealt{milgrom07b,gentile07a}, see also \citealt{bournaud07}), low surface brightness galaxies (\citealt{mcgaughdeblok,milgrom07a}) and high surface brightness galaxies (\citealt{fb05,sandnoord,mcgaugh08}) including giant ellipticals (\citealt{aftcz}).

Not only are the dynamics of the galactic systems well matched from the MOND prediction, but they all fall precisely on the Tully-Fisher relation (\citealt{mcgaugh00,mcgaugh05a}) which correlates total enclosed mass and the fourth power of the asymptotic velocity, unless they are satellites of a larger galaxy.

The only alternative theory uses cold dark matter (CDM) particles without any experimental motivation in massive, triaxial halos to provide the additional gravity needed to boost the rotation velocities of the systems with an acceleration discrepancy. There are several well-documented and as yet unresolved issues for this framework at the scales of galaxies such as the fine tuning problem of DM halos (\citealt{milgrom05,mcgaugh05b}), the cusp problem (\citealt{deblok98,mcgaughdeblok,gnedin02,gentile04,gilmore07}), the missing satellites problem (\citealt{klypin99,moore99}) and more recently tidal dwarf galaxies (\citealt{milgrom07b,gentile07a}, see also \citealt{bournaud07}). Nevertheless, it is the generally accepted model.

One might find this astounding except for the fact that it has resounding success at cosmological scales where MOND predictions are sketchy. Furthermore, clusters of galaxies require large quantities of dark matter in MOND (\citealt{sanders03,point,clowe06,aszf,sanders07,afb}). This might sound like a contradiction, but it is perfectly sensible as long as an obvious contraint is satisfied i.e. that any dark matter in MOND has a free streaming scale that is larger than galaxies and the density is low. Otherwise the dark matter would manifest itself in ordinary galaxies which would destroy the consistency of the baryonic Tully-Fisher relation (\citealt{mcgaugh00,mcgaugh05a}). 

It was postulated (\citealt{sanders03,sanders07}) that the active neutrinos (at $m_{\nu}=2eV$; very close to the experimental upper limit of 2.2eV) can provide the dark mass of clusters. Neutrinos conform to certain scaling relations in clusters such as the proportionality of the electron density in the cores of clusters to $T^{3/2}$. Nevertheless, recent studies (\citealt{point,aszf,afb}) have shown that, even under very favourable circumstances, a second species of dark matter would be necessary to explain the dynamics of the central 100kpc of clusters and groups of galaxies, which more or less rules them out the active neutrinos as good candidates.

Of course, there is no limit on the dark matter being baryonic since the necessary dark matter in clusters of galaxies is at most a few percent of the big bang nucleosynthesis (BBN) baryons, of which only 20\% or so are observed at low redshifts (\citealt{silk07,mcgaugh07}), the remainder is presumed to exist in a warm-hot intergalactic medium \cite{bregman07}. This led \cite{milgrom07d} to propose the dark matter in clusters to be a cold, molecular gas of $\sim$ Jupiter mass. These are naturally difficult to detect, but might have the serendipitous fortune of resolving the cooling flow problem (\citealt{fabian94}).

Unfortunately, even if the cluster dark matter problem were resolved, there remains the issue of cosmological dark matter. Put simply, there is compelling evidence that the Universe consists of a form of dark energy (like a cosmological constant) that forces the expansion of the Universe to accelerate at late times (\citealt{perlmutter99,schmidt98}). However, we have no idea what this dark energy is (\citealt{diaferio08a}) from a particle physics point of view, although perhaps the coincidence between $a_o$ and $cH_o$ or $c(\Lambda/3)^{1/2}$ is a strong indication (\citealt{milgrom02,milgrom08}).

With the presence of this dark energy, in order for the Universe not to expand too rapidly, there needs to be some form of matter independent of the well fixed quantity of baryons to endow the Universe with additional inertia. This additional matter serves several purposes: it allows for large structures to form more rapidly out of the expanding Universe which is shown by the matter power spectrum at large scales (\citealt{tegmark04}). Also, it drives the collapse of the photon-baryon fluid to form fluctuations in the cosmic microwave background (CMB) at well measured angular scales (\citealt{white94}) and it gives the correct distance-redshift relation (expansion history).

The underlying theory of MOND is still unknown, also it is only a classical framework, so a huge effort was made to extend MOND to the relativistic regime. In 2004, a giant leap was made in this direction by \cite{bekenstein04} and others have taken to thrashing out the predictions of other MOND inspired relativistic theories (\citealt{sanders05,skordis06,zlosnik06,zlosnik07a,zlosnik07b}). Sadly, the predictions for cosmology are not clear and there seems to be too much freedom, in contradiction to the absolute predictiveness of MOND in galaxies.

For this reason, I show here the predictions of coupling MOND with sterile neutrino dark matter using the {\it ansatz} employed by \cite{mcgaugh04} when matching the CMB with MOND i.e. that no MOND effects are present before recombination. A simple argument supporting this is that at a redshift of $z\sim$1080 the angular diameter distance to recombination $D_A=14Gpc$ and the angular scale, $\theta$, of the first (and largest) peak is $1^o$ or 0.017rad. So the physical size of the first peak $r=\theta D_A$ is $\sim$240Mpc. Since the average overdensity $\delta$ is only $1$ part in $10^5$ of the critical density $\rho_c(z)$, the typical gravities at a radius $r$ from the centre of one of these overdensities is $g=G\delta M(r)r^{-2}={4\pi \over 3} G\delta\rho_c(z)r$ where $\rho_c(z)={3H(z)^2 \over 8\pi G}$ and $H(z)^2=H_o^2[\Omega_m(1+z)^3 +\Omega_{\Lambda}$. Compiling all this gives

\beq
\protect\label{eqn:grav}
g(r)\sim \half \delta H_o^2 \left[\Omega_m(1+z)^3 +\Omega_{\Lambda}\right]r
\eeq
At $r=240Mpc$, 
\begin{eqnarray}
g=\half \cdot 10^{-5} \cdot 7.1\times 10^{-5}\left[ 0.27\cdot1081^3+0.73\right] \cdot240Mpc \sim 570a_o,\nonumber
\end{eqnarray}
where $a_o=3.6(\kms)^2pc^{-1}$ is the MOND acceleration constant. Typical accelerations so many times greater than $a_o$ are completely unaffected by MOND gravity and therefore no MOND effects should influence the CMB. However, as $z$ drops, so does $\rho_c(z)$ and thus peculiar accelerations can slide into the MOND regime. Thus, the matter power spectrum can be affected by MOND.

It is often forgotten when looking at MOND cosmology that {\it no} cold dark matter exists in MOND. Therefore, we must relax many of the constraints that are set by CDM cosmology. The most important and obvious one is that there is now a large gap in the energy-density budget since CDM is not present and it is perfectly reasonable to fill this gap with hot dark matter like neutrinos. The constraints on neutrino masses, for which cosmology is still the most stringent, must be reanalysed in light of MOND. Still, the empirical evidence from supernovae data (\citealt{schmidt98,perlmutter99}) strongly suggest the universes expansion is accelerating owed to the existence of dark energy, $\Omega_{\Lambda}$. Furthermore, the baryon budget is strongly constrained by well understood physics to be around $\Omega_bh^2\sim 0.015-0.025$ (\citealt{boesgaard85,burles01,mcgaugh04}), but this still leaves a large amount of latitude in the energy budget for DM.

Any DM, however, must be compatable with clusters of galaxies, the well understood lack of DM in galaxies in MOND and the anisotropies in the angular power spectrum of the CMB. The best candidates for such hot DM are neutrinos.
\section{Neutrinos}
\subsection{Active Neutrinos}
The three active neutrinos ($\nu_{\mu}$, $\nu_{e}$ and $\nu_{\tau}$) from the standard model of particle physics have been shown to mix between flavours by atmospheric and solar neutrino experiments (\citealt{ahmad01,ashie04}). However, the exact masses of the three active neutrinos are not yet known, only their squared mass differences. Nevertheless, the masses of all three are known to be less than 2.2eV from the Mainz-Troitz experiments (\citealt{kraus05}).

The maximum density that a neutrino species can produce after gravitational collapse is given by the Tremain-Gunn limit (\citealt{tremaine79}),
\beq
\label{eqn:tg}
{ \rho_{\nu}^{max} \over  7\times 10^{-5} \msun pc^{-3}}=\left({T \over 1keV}\right)^{1.5}\left({m_{\nu} \over 2eV}\right)^4
\eeq
for each of the three species. Thus, the density is greatly dependent on the mass of the neutrinos. However, groups and clusters of galaxies have dark matter that is much denser than can be produced by the active neutrinos even at the maximum mass of 2.2eV (\citealt{afb}). If the dark matter is indeed a neutrino like species, it must be heavier than 8eV (\citealt{afb}). There is a further problem with neutrinos at 2.2eV in that the contribution they make to the energy density of the Universe is given by 

\beq
\label{eqn:ordn}
\Omega_{\nu}=0.0205m_{\nu},
\eeq
meaning that at 2.2eV the three neutrinos make a 13.6\% contribution to the energy density of the Universe, but the maximum density is relatively low (see Eq \ref{eqn:tg}). Such a huge contribution would be easily detectable in the angular power spectrum of the fluctuations in the CMB as shown for this example in Fig~\ref{fig:sncmb}. Therefore, the active neutrinos are a very poorly motivated candidate.

\subsection{Sterile Neutrinos and the CMB}
As mentioned above, the three active neutrinos are known to have mass. Another oddity arising from this is that the active neutrinos are solely left handedly chiral, whereas all other fermions are ambidextrous. The easiest way to incorporate this into the standard model of particle physics is to introduce a right handed ``sterile neutrino". In addition, they are not simply aesthetically pleasing, the introduction of a single sterile neutrino was preferred from analysis of the Miniboone experiment by \cite{giunti08} (see also \citealt{aguilar01,maltoni07}) with a mass in the range 4eV-18eV to explain the disappearance of electron neutrinos from the beam at low energies.

In the simplest model, if the mixing angle of the sterile neutrino is low enough, then thermalisation in the early Universe can balance the abundance of the sterile and active neutrinos. In this case, the cosmological density is exactly related to their mass, as for the active ones (Eq~\ref{eqn:ordn}).

With the hypothesis that all the DM in MOND comes from a single sterile neutrino, we used the freely available CMB anisotropy code CAMB (\citealt{lewis99}) and incorporated it into a $\chi^2$ minimisation routine comparing with the data from the WMAP5 data release (\citealt{dunkley08}) and the ACBAR 2008 data release (\citealt{reichardt08}). We allowed variation of $\Omega_b$, $\Omega_{\nu_s}$, $n_s$, $d n_s/d\ln k$, $\tau$, $H_o$ and fixed the Universe to be flat meaning $\Omega_{\Lambda}=1-\Omega_b-\Omega_{\nu_s}$. 

Obviously, in this MOND inspired model there is no CDM by definition, but since the CDM model works well at producing the CMB anisotropies, we began the search by simply transferring $\Omega_{cdm}$ to $\Omega_{\nu_s}$. Furthermore, the 3 active neutrinos are taken, for simplicity, to be massless. As discussed later, it is not feasible to have a pair of very massive ($> 0.5eV$) sterile neutrinos because splitting the $\Omega_{\nu_s}$ between two or more neutrinos reduces the available mass to each neutrino thus detrimentally lowering its Tremaine-Gunn limit ($\rho_{\nu}^{max}\propto m_{\nu}^4$) and thus the gravity available to drive the collapse of the baryons prior to recombination on small scales like the third acoustic peak. This is highlighted in Fig~\ref{fig:sncmb} where the comparison is made between one sterile neutrino and two.

\begin{figure}
\includegraphics[angle=0,width=7cm]{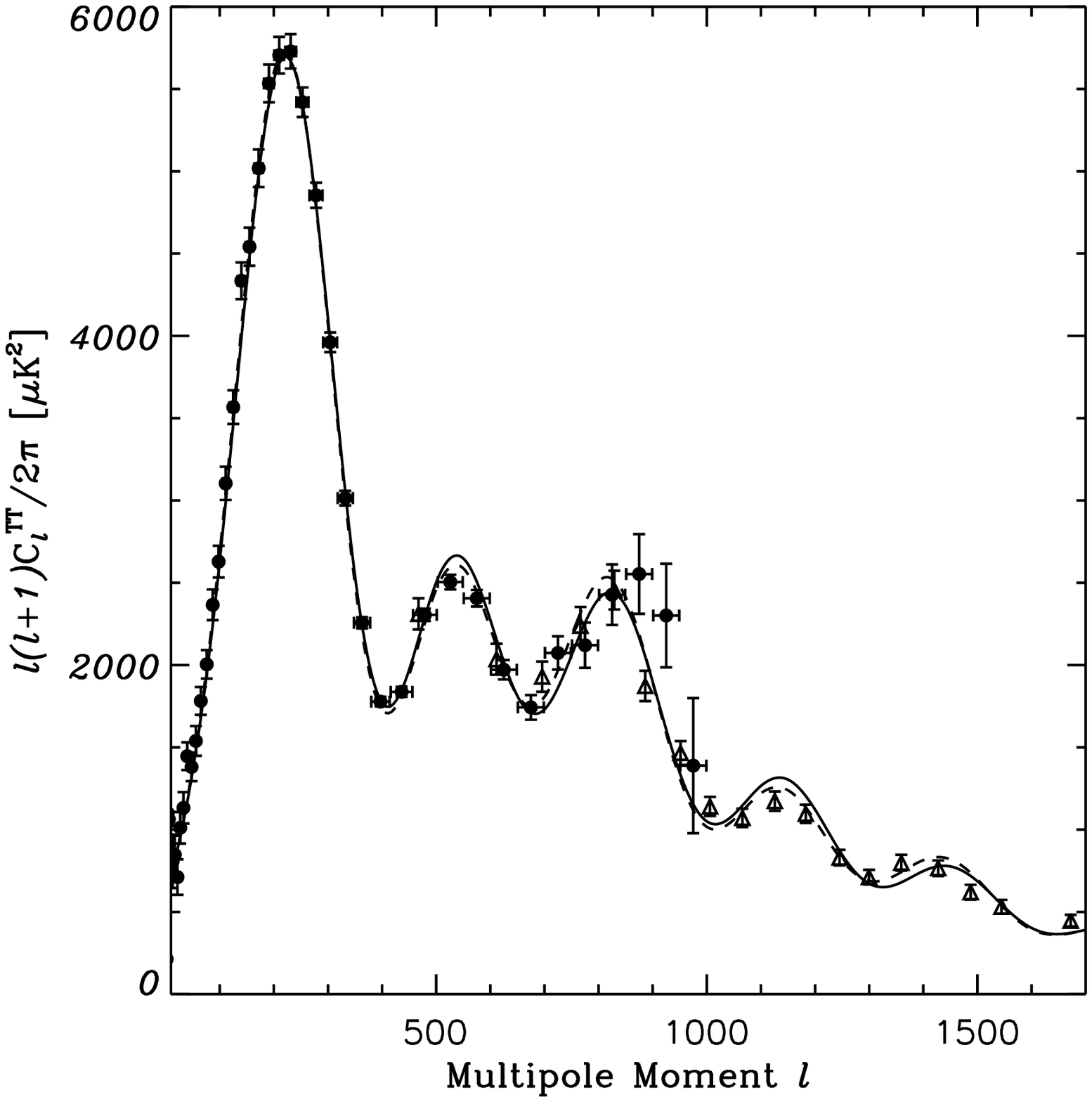}
\caption[Sterile neutrino fit to the CMB]{Shows the data of the CMB as measured by the WMAP satellite year five data release (filled circles, \citealt{dunkley08}) and the ACBAR 2008 (\citealt{reichardt08}) data release (triangles). The lines are the $\Lambda$CDM max likelihood (dashed) and the solid line is the fit with an 11eV sterile neutrino with paramters given in table 1. The $n_s$ for the $\lcdm$ model has been scaled from the quoted 0.963 in \cite{dunkley08} to 0.979 here to better match the data.}\label{fig:cmb}
\end{figure}

\begin{figure}
\includegraphics[angle=0,width=7cm]{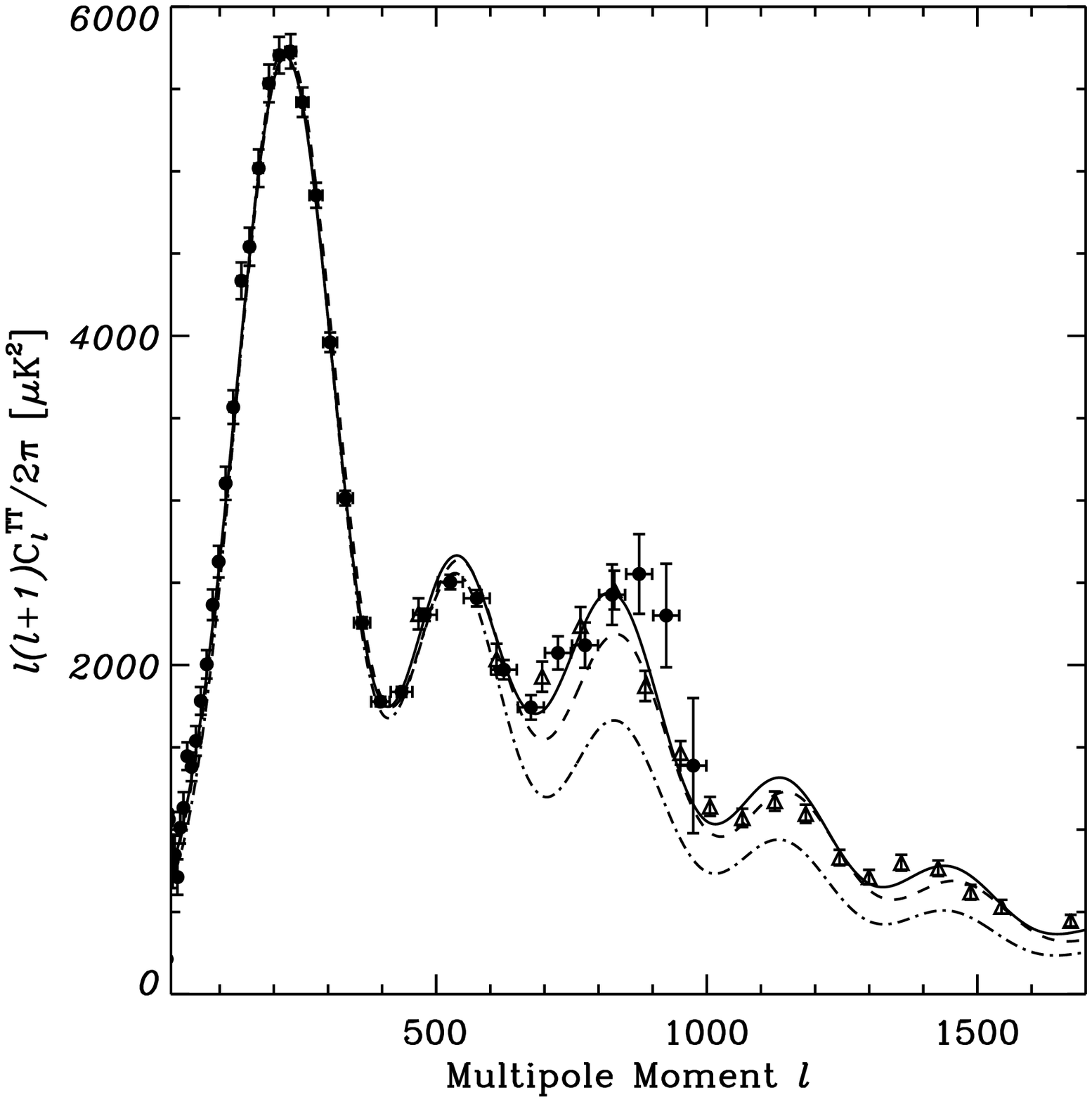}
\caption[2 Sterile neutrino fit to the CMB]{As for Fig~\ref{fig:cmb} with the solid line again the single sterile neutrino fit with parameters given in table 1, but the dashed line is the fit with 2 sterile neutrinos sharing $\Omega_{\nu_s}$ and the dotted line is with the maximum active neutrino contribution with $\Omega_{\nu}=0.136$ and $\Omega_{\Lambda}$ compensated for a flat universe. We reduced $n_s$ to 0.856 and 0.939 respectively to match the amplitude of the first acoustic peak, but the second and third peaks are badly matched because there is not enough neutrino DM density on small scales because of the Tremaine-Gunn limit.}\label{fig:sncmb}
\end{figure}

The parameters for the best fit are given in table 1 which also contains the parameters for the WMAP5 fit from \cite{dunkley08} and a comparison of the two fits are shown in Fig~\ref{fig:cmb}. All parameters are consistent with experimental bounds and are not significantly different to the $\lcdm$ model, which is sensible since the $\lcdm$ model of the CDM anisotropies is a good one.

The mass of the sterile neutrinos infered from the best fit value of $\Omega_{\nu_s}h^2=0.117$ is $m_{\nu_s}\sim11eV$. This mass range of sterile neutrino has never before been considered in the literature because it is excluded by cosmological data if we assume Newton's law are correct (\citealt{dodster,seljak06}) since they cannot influence galaxy rotation curves because they would have a free streaming scale (cf. \citealt{sanders07}) of more than $R_c=1.3 \left( {m_{\nu} \over 1eV} \right)^{-4/3} \left({V_r \over 200\kms}\right)^{1/3} = 50kpc$ in a Milky Way type galaxy, for $V_r=200\kms$. The total mass this would create within 8kpc is $\sim5\times10^9\msun$ which is about 10\% of the total mass and would actually help MOND fits to the Milky Way's rotation curve (\citealt{fb05,gentile08,mcgaugh08}).

\begin{figure}
\includegraphics[angle=0,width=7cm]{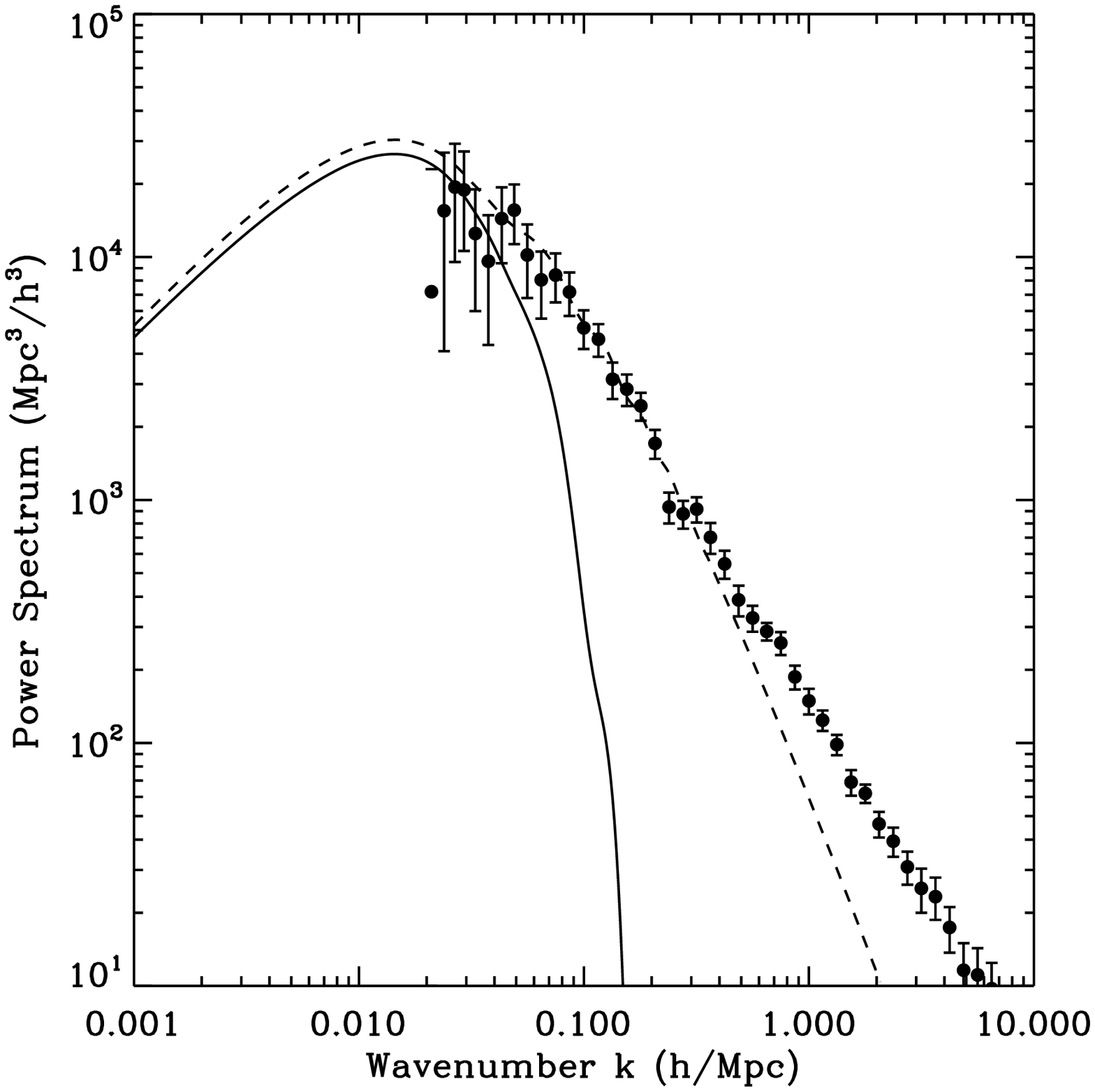}
\caption[mps]{The filled circles are the data points from the SDSS (\citealt{tegmark04}), the solid line is the single sterile neutrino model but with Newtonian instead of MONDian gravity. The dashed line is the $\lcdm$ model.}\label{fig:pk}
\end{figure}

In particular, it would have a similar contribution to the energy density as required from cold dark matter fits to the CMB ($\Omega_{\nu_s}h^2=0.117$; $\Omega_{cdm}h^2=0.108$) and leave the matter power spectrum at large scales ($>50h^{-1}$Mpc) unaltered. This is shown in Fig~\ref{fig:pk} which compares the observed matter power spectrum with that predicted by the sterile neutrino model here, but with Newtonian instead of MONDian gravity. At scales smaller than $\sim50h^{-1}$Mpc the computed power spectrum drops many orders of magnitude below the observed one.

Qualitatively, this discrepancy is owed to the fact that structures on these small scales have formed with the assistance of MONDian gravity. For instance, following the argument of Eq~\ref{eqn:grav}, the redshift by which scales as large as $50h^{-1}$Mpc are deep in the MOND regime (i.e. $g\sim {a_o \over 10}$) is roughly

\beq
\protect\label{eqn:red}
z\sim \left[ {2 g \over \delta H_o^2 \Omega_m r } \right]^{1/3},
\eeq
which for 70Mpc is $z \approx 100$. Certainly many authors (\citealt{sanders08,nusser02,knebe04}) have shown that structures can form very quickly in MOND even without CDM and galaxy size objects can be in place as early as $z \approx 10$.

The tools to perform the full matter power spectrum analysis are currently not available for MOND (nor standard dynamics), since they crucially depend on hydrodynamics. Assuming that including the modified dynamics enables a match to the matter power spectrum at all scales, the only conceivable ways of distinguishing between MOND and $\lcdm$ (if missing satellites, the lack of cusps in DM halos and tidal dwarf galaxies are ignored) is in the complex modelling of galaxy formation, or the unambiguous detection of the hot or cold DM particles.

\begin{table} 
\begin{tabular}{|c|c|c|c|c|}
\hline
Parameter & Single $\nu_s$ & $\lcdm$ & Active & Two $\nu_s$\\
\hline
$H_o$ & 71.5 &  72.4  &  71.5  &  71.5 \\
$100\Omega_{b}h^2$ & 2.4 &  2.27  &  2.4  &  2.4 \\
$\Omega_{\nu}h^2$ & 0.117 &  0.0  &  0.695  &  0.117 \\
$\Omega_{cdm}h^2$ & 0.0 &  0.108  &  0.0  &  0.0 \\
$n_s$ & 0.965 &  0.979  &  0.856  &  0.939 \\
No. massless $\nu$ & 3 &  3  &  0  &  3 \\
No. massive $\nu$ & 1 &  0  &  3  &  2 \\
\hline
\end{tabular} 
\medskip 
\caption{List of parameters used in the figures. The $\lcdm$ numbers come from Dunkley et al. (2008) but $n_s$ has been scaled from the quoted 0.963 to 0.979 for a better match to the data.} 
\end{table}

\section{Discussion}
A single massive sterile neutrino appears consistent with the current level of precision in the measurements of the CMB anisotropies. It is also consistent with the matter power spectrum at large scales ($>50h^{-1}$Mpc) and is able to clump together with densities surpassing the maximum density of the DM in groups and clusters of galaxies where MOND requires dark matter of some form. As discussed in \cite{afb} there appears to be a scale at which MOND begins to poorly describe the dynamics of astrophysical systems. This is highlighted by \cite{romanowsky03,milgrom03,aftcz,osullivan08} which show that no dark matter is necessary to explain the detailed dynamics of relatively low mass groups of galaxies and systems smaller. This is expected for sterile neutrino dark matter because it would have a free streaming length greater significantly larger than a typical galaxy ($\sim$50kpc for the Milky Way. However, just as numerical simulations of clusters of cold dark matter were necessary to show that the CDM halos are a poor match to observed galaxies (\citealt{deblok98,mcgaughdeblok,gnedin02,gentile04,gilmore07}), the equilibrium distribution of the sterile neutrino DM must be checked to be consistent with groups and clusters of galaxies (see \citealt{sanders07}).

On the other hand, the three active neutrinos should probably have masses well below 0.5eV. Otherwise it will become difficult to match the CMB power spectrum because the angular scale of the peaks prefers $\Omega_{\nu}h^2=0.117$ while $\Omega_{\nu}\propto m_{\nu}$. Increasing the mass of another neutrino reduces the mass of the sterile neutrino and the amplitude of the third peak of the CMB diminishes due to the rapidly decreasing maximum density ($\rho_{\nu}^{max}\propto m_{\nu}^4$).

Certain analyses of neutrino mixing experiments seem to require an additional, sterile neutrino with a mass in the range $4eV<m_{\nu_s}<18eV$. Here I took the {\it ansatz} that there is a fourth, sterile neutrino of 11eV mass and that MOND effects are not important at cosmological scales. I showed that its contribution to the dynamics of galaxies would be negligible, but that it could solve all problems MOND has with the dynamics of clusters of galaxies and it can match the angular power spectrum of the CMB. The matter power spectrum needs to be recalculated because MOND gravity is crucial to the formation of these smaller structures and because of the increased dominance of baryons at these scales over DM, hydrodynamics cannot be avoided as in CDM simulations. If experiments can indeed pinpoint the existence of a sterile neutrino with mass $\sim$11eV this would be a significant advance for the Modified Newtonian Dynamics.

Even if collider experiments detect a CDM candidate with mass of 300GeV, this will give us virtually no information about the cosmological abundance and therefore brings us no closer to solving the dark matter problem. The great thing about sterile neutrinos is that if we can find the mass from laboratory experiments then this effectively fixes the cosmological abundance AND the contribution the neutrinos can make to clusters of galaxies can be strictly constrained. Henceforth, it would be possible to run structure formation simulations in MOND with all the ingredients.

\section{acknowledgments}GWA's research was supported by an STFC travel grant and an STFC studentship.


\label{lastpage}

\end{document}